\documentclass[aps,prl,twocolumn,superscriptaddress]{revtex4}
\usepackage{graphicx}

\begin{document}

\title{One-Dimensional Self-organization and Nonequilibrium Phase Transition
in a Hamiltonian System}
\author{Jiao Wang}
\affiliation{Department of Physics and Institute of Theoretical Physics
and Astrophysics, Xiamen University, Xiamen 361005, Fujian, China}
\author{Giulio Casati}
\affiliation{Center for Nonlinear and Complex Systems,
Universit\`a degli Studi dell'Insubria, via Valleggio 11, 22100 Como, Italy}
\affiliation{International Institute of Physics, Federal University of
Rio Grande do Norte, Campus Universit\'ario - Lagoa Nova, CP. 1613, Natal,
Rio Grande Do Norte 59078-970, Brazil}

\date{\today}

\begin{abstract}
Self-organization and nonequilibrium phase transitions are well known to occur in
two- and three- dimensional dissipative systems. Here, instead, we provide numerical
evidence that these phenomena also occur in a one-dimensional Hamiltonian system.
To this end, we calculate the heat conductivity by coupling the two ends of our
system to two heat baths at different temperatures. It is found that when the
temperature difference is smaller than a critical value, the heat conductivity
increases with the system size in power law with an exponent considerably smaller
than 1. However, as the temperature difference exceeds the critical value, the
system's behavior undergoes a transition and the heat conductivity tends to diverge
linearly with the system size. Correspondingly an ordered structure emerges. These
findings suggest a new direction for exploring the transport problems in one
dimension.
\end{abstract}


\maketitle

Self-organization phenomena~\cite{SO} are ubiquitous in nature and human
society. Among others, nonequilibrium phase transitions (NPTs) are a familiar
route leading to collective organized behavior~\cite{NPT}. A well-known example
is the Rayleigh-B\'enard convection~\cite{RB} observed in a horizontal fluid
layer heated from below. When the temperature difference between the heater and
the thermal sink (set on the fluid layer) is low, the fluid layer is static and
the heat is transferred uniformly from bottom to top. But when the temperature
difference reaches a critical value, the microscopic random movement spontaneously
becomes ordered on a macroscopic level. The fluid  spontaneously organizes into
a regular pattern of convection cells and the heat flux is dramatically increased
compared to the molecular heat conduction. Rayleigh-B\'enard convection indicates
that a physical system may have the ability to radically change its structure to
adaptively respond to strong thermodynamic driving.

These phenomena of self-organization and NPTs are known to take place in two and
three dimensions. In the one-dimensional (1D) case instead, despite extensive analytical,
numerical, and experimental investigations, they have not been observed yet in any
physical systems~\cite{Lepri14}. A class of systems where 1D NPTs are intensively
studied is given by hopping models in which particles hop from one site to another
on a 1D lattice with a prescribed stochastic dynamic~\cite{Evans00}. These models
have been found useful in a wide range of applications such as directed percolation,
growing interfaces, traffic flows, and so on. They are not suitable, however, to
study problems where physical quantities like energy and momentum are involved.

On the other hand, in the past few decades the study of 1D transport problems
has attracted considerable interest~\cite{Lepri03-16, Dhar08, Lee-Dadswell}. A
remarkable result states that for a general 1D momentum conserving system, the
heat conductivity diverges with the system size in a power law and the diverging
exponent may take a universal value under certain conditions~\cite{Lee-Dadswell,
Narayan, Lepri0607, Beijeren12, Spohn13}. It is worth noting that a common basis
for different theoretical approaches is the linear response theory that is valid
near equilibrium. Therefore, NPTs, which usually occurs far from equilibrium,
have been implicitly excluded.

In this Letter we show that NPT and self-organization can take place in a 1D
Hamiltonian system. Our result also suggests the need to extend the study of
1D transport problems beyond the linear response regime. In principle, this
might reveal new mechanisms for energy, momentum, and matter transport.

Our model is a combination of the classical version of the Lieb-Liniger model
\cite{LL63-1,LL63-2} and the two-mass gas model introduced in Ref.~\cite{Casati86}.
It consists of $N$ point particles with short-range interaction, confined
in a 1D box of length $L$. The Hamiltonian is
\begin{equation}
H=\sum_{i} \frac{p_i^2}{2m_{i}}+\sum_{i<j}V(x_{i}-x_{j}),
\end{equation}
where $p_i$, $m_i$, and $x_i$ are, respectively, the momentum, mass, and position
of the $i$th particle. The mass of a particle is either $\mu_1$ or $\mu_2$, and
the potential $V(x)$ is a step function: $V(x)=h$ for $x \le |r|$ and $V(x)=0$
otherwise, with $h\ge 0$ being the potential barrier. In this work we consider
the limit case $r\to 0$. This case has been considered in Ref.~\cite{Kurchan02}
in order to study the Soret effect which basically consists in a formation of
a density gradient when a binary fluid is subjected to a temperature gradient.

According to the Hamiltonian Eq.~(1), all particles move freely, and when
two particles meet, either they pass through each other without changing their
velocities or they collide elastically, depending on their total energy. More
precisely, if their total energy in the frame of the center of mass is larger
than $h$, or equivalently, if their relative velocity (speed) is larger than a
certain value, i.e., $|v_i-v_{j}|>\sqrt{2h(m_{i}+m_{j}) /(m_{i}m_{j})}$, then
the two particles pass through each other; otherwise they collide elastically.
Note that the interaction between particles conserves their energy and momentum,
and that the limiting case $h=0$ corresponds to particles moving freely and
independently while $h=\infty$ corresponds to the standard 1D hard point
gas~\cite{Casati86}. Also note that for a finite $h$ the potential $V(x)$
is symmetric (with respect to $x=0$), but for $h=\infty$, due to the fact that
particles do not cross each other, the potential function is in effect asymmetric.

To compute the heat conductivity of the system, we couple its two ends to
two heat baths at temperature $T_L=T+\Delta$ and $T_R=T-\Delta$, respectively.
When the first (last) particle of mass $\mu_{i}$ hits  the left (right)
boundary of the box, it is reflected back with a new velocity $v$ given
by the probability distribution~\cite{Bath}:
\begin{equation}
P_{L,R}(v)=\frac{|v|\mu_{i}}{k_BT_{L,R}}\exp{(-\frac{v^2 \mu_{i}}{2k_B T_{L,R}})},
\end{equation}
where $k_B$ is the Boltzmann constant. The heat conductivity is then evaluated
as $\kappa= jL/(T_L-T_R)$, with $j$ being the thermal current. Since in our
system the heat conductivity follows the scaling $\kappa(cT, c\Delta, ch)
=\sqrt{c}\kappa(T,\Delta,h)$, we fix $T$ and investigate the dependence of
$\kappa$ on $\Delta$ and $h$. For details of numerical simulation, see Ref.
\cite{Chen14}. For all data points shown in the figures the errors are at most
$3\%$, and as the error bars are smaller than the symbols, they are omitted.

First, we have checked that the results do not qualitatively depend on the
particular temperature $T$ and mass values. Therefore, we choose $k_B=1$,
$T=5$, $N/2$ particles with mass  $\mu_1=1$ and $N/2$ particles with mass
$\mu_2=2$. We fix $L=N$ so that, on average, there is one particle in a unit
volume (length). We have also verified that the results do not depend on
initial conditions; in particular, and quite importantly, as long as $h\ne
\infty$, they do not depend on the initial particle-mass configuration either.

For the special, limiting case $h=0$, particles move freely and independently.
We therefore have
\begin{equation}
\kappa=N\sqrt{\frac{k_B^3}{2\pi}}(\frac{1}{\sqrt{\mu_1}}+\frac{1}{\sqrt{\mu_2}})/
(\frac{1}{\sqrt{T_L}}+\frac{1}{\sqrt{T_R}}).
\end{equation}
Up to the first order in $\Delta/T$,  $\kappa$ does not depend on $\Delta$, while
the second-order correction is only $1.5\%$ for $\Delta/T=0.2$. Hence $\Delta=1$
(with $T=5$) can be considered inside the linear response regime (see Fig.~1).
For the opposite limiting case $h=\infty$ numerical simulations provide empirical
evidence that $\Delta=1$ is also within the linear response regime. In addition,
based on the linear response theory it is predicted that $\kappa\sim N^{1/3}$ in
the thermodynamic limit~\cite{Narayan, Lepri0607, Beijeren12, Spohn13}. This result
is supported by our numerical simulations as well, which give $\kappa\sim N^{0.30}$
for $10^3<N<4\times 10^4$. (More simulation results for other mass values of
$\mu_1$ and $\mu_2$ and a bigger temperature difference $\Delta=3$ can be found
in Ref.~\cite{Yang}.) In summary, for both limiting cases $h=0$ and $h=\infty$
there is a linear response regime up to $\Delta=1$ in which the heat conductivity
does not depend on the temperature difference $\Delta$ significantly.

\begin{figure}[!]
\includegraphics[width=8.5cm]{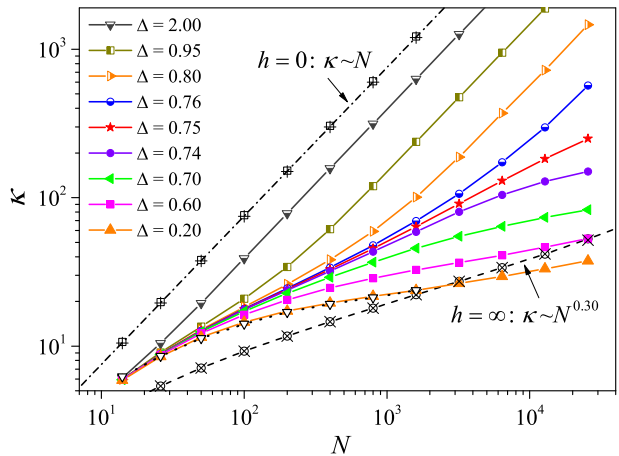}
\vskip-0.1cm
\caption{The heat conductivity as a function of the system
size for $h=10$ and for several temperature difference values $\Delta>0$. As
a comparison, the black open triangles connected by the dotted line segments
are for the heat conductivity computed in the equilibrium state (corresponding
to $\Delta=0$) by using the Green-Kubo formula~\cite{Dhar09}. For reference,
we show also the case $h=0$ for $\Delta=0.2$ (open squares) and $\Delta=1.0$
(plus signs) and the case $h=\infty$ for $\Delta=0.2$ (open circles) and
$\Delta=1.0$ (crosses). The straight dot-dashed line is the result given
by Eq.~(3) up to the first order of $\Delta/T$. }
\vskip-0.3cm
\end{figure}

In clear contrast, for a finite but nonzero value of $h$, the dependence of
the heat conductivity on the system size $N$ and on the temperature difference
$\Delta$ is completely different. A typical result is shown  in Fig.~1 for $h=10$.
It can be seen that the dependence of $\kappa$ on $N$ dramatically depends on the
temperature difference $\Delta$ and the most striking feature is the existence of
a critical value of $\Delta$, $\Delta_{cr}=0.75$, that divides the $N$ dependence
of $\kappa$ into two separate classes: For $\Delta>\Delta_{cr}$, $\kappa$ tends
to increase linearly with $N$, while for $\Delta<\Delta_{cr}$, $\kappa$ tends to
increase with $N$ as $\kappa\sim N^{\alpha}$, with $\alpha \approx 0.17$. At the
special value $\Delta$=$\Delta_{cr}$, $\kappa$ tends to increase with $N$ in power
law as well but with an exponent of $\alpha \approx 0.50$ instead (see red stars
in Fig.~1). It is worth noting that, since the numerical values of $\alpha$ are
evaluated based on the simulation results at finite system sizes, one should be
very careful in attempting to extrapolate them to the thermodynamic limit, which
is, as usual, a delicate question~\cite{Lepri03-16}.

\begin{figure*}[!]
\includegraphics[width=17.9cm]{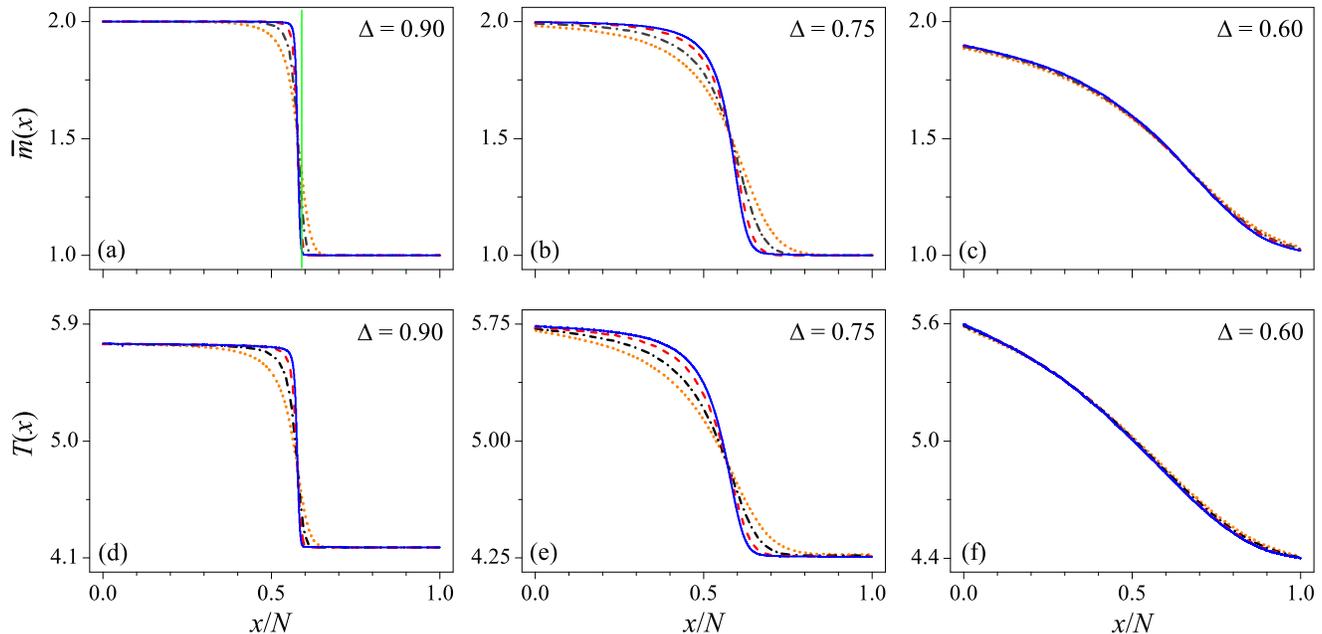}
\vskip-0.2cm
\caption{(a)-(c) The average mass of particles at position $x$,
at the stationary state, for the same value $h=10$ as in Fig.~1. (a) $\Delta =0.9
>\Delta_{cr}$, (b) $\Delta=0.75=\Delta_{cr}$, and (c) $\Delta =0.6<\Delta_{cr}$,
respectively. The corresponding temperature profile is shown in (d)-(f). In all
six panels, the orange dotted, the gray dash-dotted, the red dashed, and the
blue solid line are, respectively, for the system size $N=1600$, $3200$, $6400$,
and $12800$. The green vertical line in (a) indicates the position of the interface
of the two species of particles if they are assumed to separate entirely.}
\end{figure*}

We conjecture that $\Delta_{cr}$ is a critical point based on the following
two observations. First, the heat conductivity as a function of the system size
depends on $\Delta$ very sensitively in the neighborhood of $\Delta_{cr}$; e.g.,
changing $\Delta$ by just 0.01 away from $\Delta_{cr}$ causes a transition
from one class to the other of the dependence $\kappa$ versus $N$ (see the two
nearest neighboring curves next to $\Delta_{cr}$ in Fig.~1). This is a strong
signal of critical phenomena. Second, the structure of the system at the stationary
state is qualitatively different for $\Delta>\Delta_{cr}$ and $\Delta<\Delta_{cr}$.
This is illustrated in Fig.~2, where we plot $\overline m(x)$, i.e., the average
mass of the particles passing a given position $x$. It is interesting to note that
for $\Delta>\Delta_{cr}$ [see Fig.~2(a)], the system is characterized by an ordered
``sandwich'' structure: while heavy particles of mass $\mu_2$ aggregate at the hot
(left) end, light particles of mass $\mu_1$ aggregate at the cold (right) end, and
in between there is a transition layer. This layer is characterized by a steep linear
center whose width does $not$ change as the system size $N$ increases. Therefore, in
the variable $x/N$, the width of the transition layer shrinks to zero as $N$ increases
[see Fig.~2(a)]. The vertical line at $x/N= T_L/(T_L+T_R)=0.59$ in Fig.~2(a) corresponds
to the ideal situation of complete separation of the two species of particles with
different masses.

It is interesting to investigate how fast the average particle mass at the
left and right ends of the system converge to $\mu_2$ and $\mu_1$ respectively,
for $\Delta>\Delta_{cr}$. Numerical results shown in Fig.~3 (for the left end)
indicate the scaling $\mu_2-\overline m(0)\sim N^{-1.5}$.

\begin{figure}[!]
\vskip-0.05cm
\includegraphics[width=8.6cm]{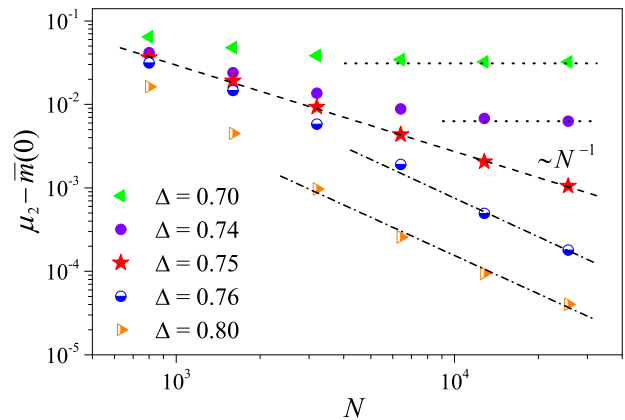}
\vskip-0.1cm
\caption{The average particle mass $\overline m(0)$ at the
leftmost end of the system for $h=10$. The dotted, dashed, and dot-dashed
lines indicate, respectively, the scaling $\sim N^{0}$, $\sim N^{-1}$, and
$\sim N^{-1.5}$.}
\vskip-0.3cm
\end{figure}

The ordered sandwich structure explains satisfactorily the asymptotic scaling
$\kappa\sim N$ for $\Delta>\Delta_{cr}$, because increasing the number of particles,
or, equivalently, the size of the system, basically only causes the two ending
aggregating segments to increase their sizes, but the thermal resistance which
is mainly due to the transition layer does not change significantly. As a consequence
the heat current does not change significantly either. Accordingly, the temperature
profile $T(x)$ also exhibits a sandwich structure: since the particles in the two
ending segments are almost identical, no noticeable internal temperature gradient
can appear; a temperature gradient is expected only in the transient layer, as is
indeed confirmed by the results in Fig.~2(d).

For $\Delta<\Delta_{cr}$, the sandwich structure vanishes: the function $\overline
m(x)$ for different system sizes agrees with each other almost perfectly upon scaling
$x \to x/N$ [Fig.~2(c)] and so does $T(x)$ [Fig.~2(f)]. Quite obviously, as $\Delta$
decreases, the functions $\overline m(x)$ and $T(x)$ will become more and more
homogeneous till completely uniform at $\Delta=0$. Such a distinctive feature
convincingly demonstrates that our system has two different nonequilibrium
phases.

At $\Delta=\Delta_{cr}$ [see Fig.~2(b)], both the sandwich and the scaling-invariance
features are lost. Here, as $N$ increases, $\overline m(x)$ at the two ends of the
system still tends to $\mu_2$ and $\mu_1$. However, quite strikingly, a good and
different scaling $\mu_2-\overline m (0)\sim N^{-1}$ emerges over a wide system size
range (Fig.~3), supporting again that $\Delta = \Delta_{cr}$ is a critical point.

\begin{figure}[!]
\vskip-0.05cm
\includegraphics[width=8.45cm]{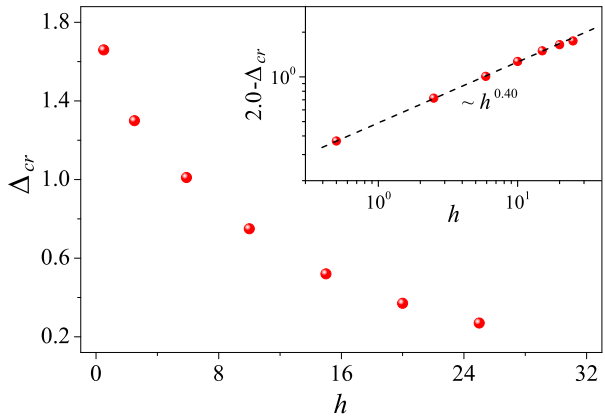}
\vskip-0.1cm
\caption{The dependence of the critical temperature difference $\Delta_{cr}$ on
the interaction parameter $h$. Inset: Same as the main panel but in log-log scale.}
\vskip-0.3cm
\end{figure}

So far we have investigated the case  $h=10$. The same critical phenomenon
takes place for any finite value of $h>0$ we have considered. Quite obviously, the
numerical value of $\Delta_{cr}$ (as well as the exponent $\alpha$ for $\Delta=
\Delta_{cr}$) depends on $h$. This is illustrated in Fig.~4, where it is seen that
$\Delta_{cr}$ decreases as $h$ increases. This fact may appear counterintuitive at
first sight and call into question the mechanism according to which heavy (light)
particles tend to reside in the hot (cold) end. The simplest case of two particles,
one light and one heavy, may shed some light. In this case, by taking into
consideration the dynamics and the heat bath model [see Eq.~(2)], it can be shown
straightforwardly that the probability for the two particles to cross each other is
higher when the light particle is at the hot end than vice versa. This is because
the relative velocity (speed) of the two particles is larger when the light particle
is at the hot end. In addition, it can be shown further that this asymmetry becomes
enhanced as $\Delta$ increases at a fixed $h$, or as $h$ increases at a fixed
$\Delta$. This mechanism should also work in a larger system.

Quite clearly, NPT is not present in the limiting cases $h=0$ and $h=\infty$. In
the former case particles move independently and the system is integrable, while
in the latter case the particles are not allowed to change their order; hence, the
mass configuration is determined by the initial setting.

Finally, we have considered so far the case in which the ratio of the two species
of particles with different masses is unity. We remark that the NPT properties
do not depend on this ratio sensitively. For example, when this ratio is decreased
to 2/8 (light to heavy) the value of $\Delta_{cr}$ is still $0.75$ within numerical
accuracy.

To summarize, NPT and self-organization have been shown to take place in
a 1D Hamiltonian system. It is found that the system can adjust its structure
adaptively to respond to the external thermal driving. In particular, for
$\Delta> \Delta_{cr}$, an ordered sandwich structure appears, leading to
the asymptotic linear dependence $\kappa\sim N$. For $\Delta< \Delta_{cr}$,
as the system size increases, the structure of the system tends to be scaling
invariant and the heat conductivity tends to increase as the system size in
power law with the exponent being considerably smaller than 1.

Importantly, in our system a linear response regime comparable to the
cases  $h=0$ and $h=\infty$ does not exist. [To be precise, our results do
not exclude the existence of a linear response regime in a much narrower range
of $\Delta$. For example, for $h=10$, the case we have mainly focused on, a
linear response regime may exist within $\Delta<0.2$ (see Fig.~1).] The heat
conductivity measured in the nonequilibrium setup depends on $\Delta$ and, hence,
is different from that obtained with the Green-Kubo formula by integrating
the heat current autocorrelation function in the equilibrium state. The
behavior of our model is, therefore, another confirmation that
``nonequilibrium is different''~\cite{Dorfman15, Politi11} and
calls for a new theoretical approach in dealing with transport problems
in systems of similar type.

As it has been shown here, NPT in our 1D system not only manifests itself in the
heat conduction behavior but also in the mass distribution. A related significant
problem is how NPT and self-organization may influence the 1D matter transport
in a physical system. Important practical examples include molecules
and micelles translated along tubulous filaments~\cite{Kral13}. Of particular
interest is to investigate the relevance for the coupled heat and particle current which
determines thermoelectric power generation and refrigeration~\cite{Benenti13, Benenti16}.
This study is in progress.

We acknowledge support by NSFC (Grants No. 11535011, No. 11335006, and No. 11275159),
by MIUR-PRIN, and by the CINECA project Nanostructures for Heat Management
and Thermoelectric Energy Conversion.

\end{document}